\documentclass{elsart} %elsart3p}

\usepackage{amssymb,graphicx}

\def\fig#1{Fig.~\ref{#1}}
\def\eq#1{Eq.~(\ref{#1})}

\newcommand{\s} {\ensuremath{\sqrt{s}}}
\def\gev{\mbox{~GeV}}
\def\gevc{\mbox{~GeV/$c$}}

\def\mevc{\mbox{~MeV/$c$}}
\def\pt#1{\ensuremath{p_{T\rm #1}}} 
\def\ptkv#1{\ensuremath{p^2_{T\rm #1}}} 
\def\vpt#1{\ensuremath{\vec{p}_{T\rm #1}}} 

\def\eg{{\it e.g}}
\def\la{\left< }
\def\ra{\right> }
\def\mean#1{\ensuremath{\la#1\ra}}

\def\meankv#1{\ensuremath{\la#1^2\ra}}
\def\rms#1{\meankv{#1}}
\def\sqrtrms#1{\ensuremath{\sqrt{\meankv{#1}}}}

\def\kt#1{\ensuremath{k_{T\rm #1}}}

\def\jt#1{\ensuremath{j_{T\rm #1}}}

\def\ptq#1{\ensuremath{\hat{p}_{T\rm #1}}} 
\def\ptqkv#1{\ensuremath{\hat{p}^2_{T\rm #1}}}

\newcommand{\dau} {\ensuremath{d+Au}}
\newcommand{\pp} {\ensuremath{p+p}}
\newcommand{\ee} {\ensuremath{e^+ + e^-}}
\newcommand{\piz} {\ensuremath{\pi^0}}
\newcommand{\pout} {\ensuremath{p_{\rm out}}}

\newcommand{\zt} {\ensuremath{z_{\rm t}}}
\newcommand{\mzt} {\mean{\zt}}

\newcommand{\xh} {\ensuremath{x_{\rm h}}}
\newcommand{\xhq} {\ensuremath{\hat{x}_{\rm h}}}

\newcommand{\xzkt} {\ensuremath{ \xhq^{-1}\mean{\zt}\sqrtrms{\kt{}} }}

\def\ktfinal{\ensuremath{ \sqrtrms{\kt{}} = 2.68 \pm 0.07(\rm stat) \pm 0.34(\rm sys) \gevc }}

\newcommand{\D} {\ensuremath{D^q_\pi}}
\newcommand{\sq}{\ensuremath{\Sigma_q(\hat{p}_T)}}

\def\pn{\ensuremath{\hat{p}_{\rm n}}} 
\def\pnkv{\ensuremath{\hat{p}^2_{\rm n}}}

\begin{document}

\begin{frontmatter}

\title{Jet properties from di-hadron correlations in $p+p$ collisions at $\sqrt{s}$=200 \gev}

\author{Jan Rak for the PHENIX collaboration}

\address{Department of Physics\\
P.O.Box 35 (YFL) Jyv\"askyl\"a\\
FI-40014 University of Jyv\"askyl\"a \\
Finland, EUROPE}

\begin{abstract} %%%%%%%%%%%%%%%%%%
An analysis of high \pt{}\ hadron spectra associated with high \pt{} \piz\  particles in \pp\ collisions at \s=200 \gev\ is presented. The shape of the azimuthal angular correlation is used to determine the  value of partonic intrinsic momentum \ktfinal. The effect of \kt{}-smearing of inclusive \piz\ cross section is discussed.
\end{abstract} %%%%%%%%%%%%%%%%%%

\begin{keyword}
jets, parton, intrinsic momentum, \kt{}-smearing, fragmentation
\PACS 25.75.Dw
\end{keyword} 
\end{frontmatter}

% main text
\section{Introduction}
\label{sec:intro}
An observation of similar suppression patter pattern of heavy and light quarks at RHIC
(see \eg\ \cite{Adler:2005xv}) initiated a discussion on detailed mechanism of parton energy loss in the exited nuclear medium \cite{Wicks:2005gt,Armesto:2005iq}. In order to explain the relatively strong suppression patterns of high \pt{}\ electrons from semi-leptonic $D$ and $B$ mesons decays,  collisional energy loss has to be  taken into account. However, this type of parton interaction with the exited nuclear medium should lead to the broadening of the away-side correlation peak \cite{Vitev:2005_LargeAngleCorrel}. Recent measurements of  high \pt{} trigger-associated particles do not indicate such a broadening \cite{Adams:2006yt}
although for the trigger transverse momentum $\pt{t}\sim 3$ GeV/c with lower \pt{}\ associated particles, significant broadening is observed \cite{Adler:2005ee}. It is evident that the measurement of jet fragmentation properties and intrinsic parton transverse momentum \kt{}\ for \pp\ collisions is vital for the understanding of pQCD phenomena and it also provides an important baseline for comparison to the results in heavy ion collisions.

In the center-of-mass of the hard collision  the two scattered partons propagate  back-to-back with the opposite momenta.  In the center-of-mass of the \pp\ collision a finite net
\pt{pair}\ results in an acoplanarity and momentum imbalance of the di-jet
which is measured as \kt{}. In the case of the Drell-Yan proccess the di-lepton net pair momentum reflects directly the magnitudes of \kt{}\ vectors of $q\bar{q}$ parton pair in the annihilation process. In the case of di-hadron correlation the jet fragmentation process is involved and it is commonly accepted to characterize the di-hadron acoplanarity by quantity $\
 = {| \vpt{t}\times \vpt{a}|  / \ptkv{t}}$, the transverse momentum component of the away-side particle \vpt{a}\ perpendicular to trigger particle \vpt{t} (for more details see \cite{longPRC:2006sc}).
It is believed that the origin of \kt{}\ is three fold:
I. Fermi motion induced by the finite transverse size of the nucleon wave-function.
II. Soft gluon showering (see \eg\ \cite{Werner_resummation}).
III. Hard next to leading order radiation.

The contribution to the \kt{}\ from the Fermi motion is $\approx\hbar/r_N\cong$300 \mevc\  where $r_N$, the transverse size of the nucleon wave-function, is of order of 1 fm. However, the measured values of \mean{\kt{}}\ over the broad range  of center-of-mass energies are as large as 5 \gevc\ \cite{Apanasevich_kt_E609}. 
\begin{figure}[h]
\begin{center}
  	\resizebox{7cm}{!}{\includegraphics{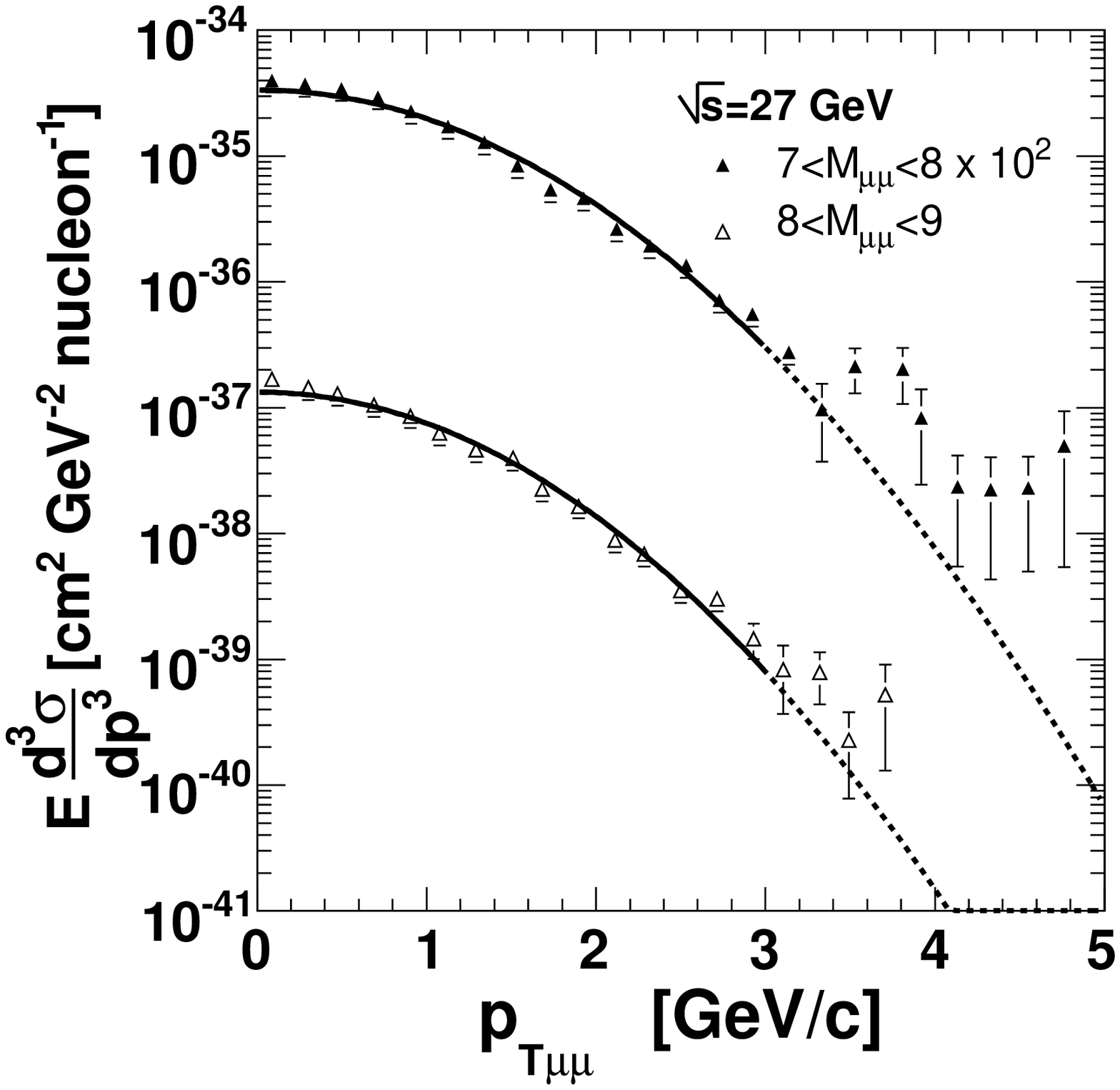}}
  	\resizebox{6cm}{6.3cm}{\includegraphics{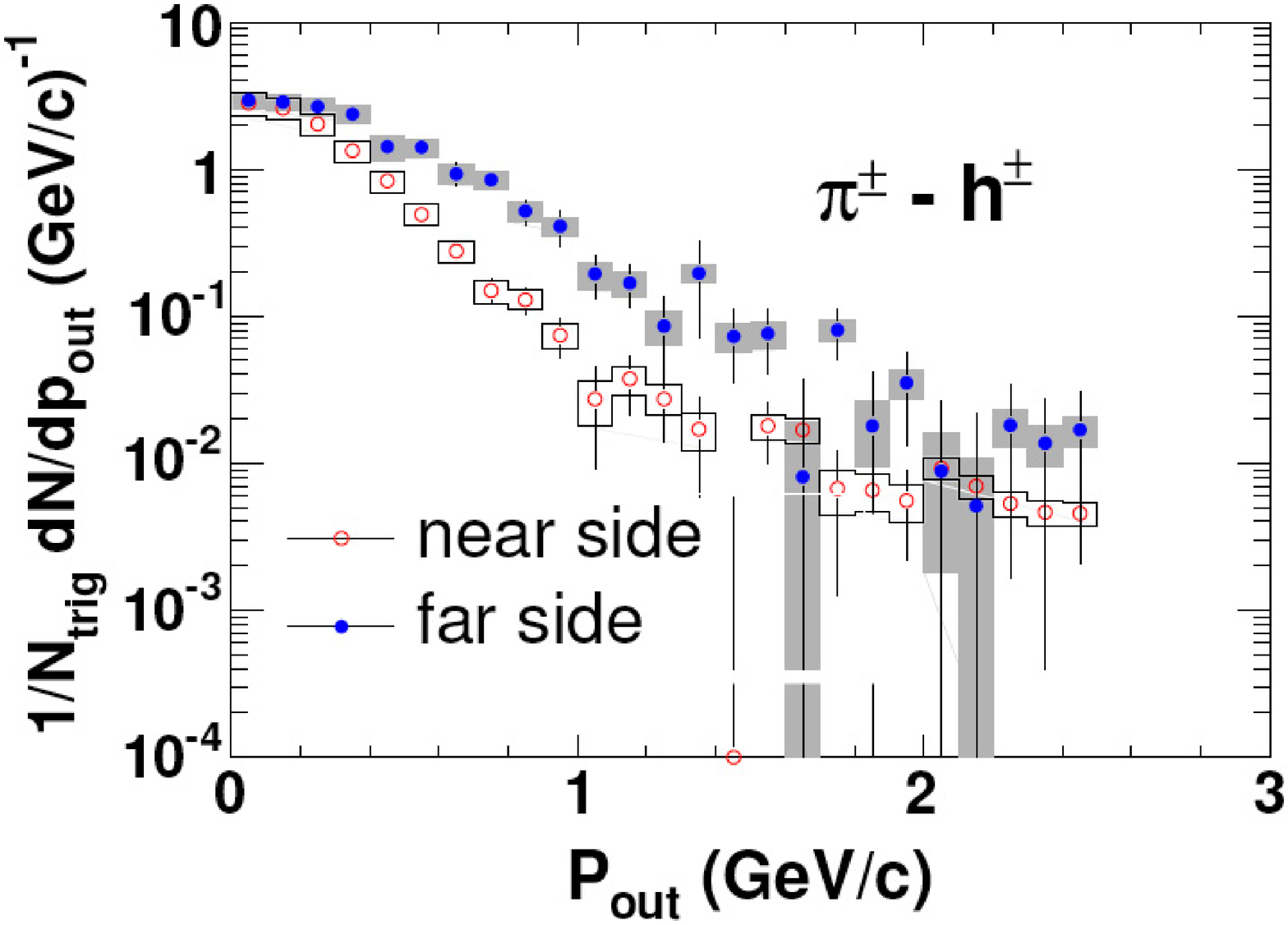}}
	\caption{{\sl Left panel:} Muon pair transverse momentum distribution for the 
	invariant mass in $7<M_{\mu\mu}<8$ (solid triangles) and $8<M_{\mu\mu}<9$ \gevc\
	 (empty triangles) range measured  in the fixed target experiment ($p_{beam}$=400\gevc ) 
	 \cite{Ito:1980ev}.  The solid lines represent the Gaussian fit in the 
	 $0<\pt{\mu\mu}<3$~\gevc. Dashed line represent the extrapolation of the fit function up to 
	 \pt{\mu\mu}=5~\gevc~region.
	 {\sl Right panel:} \pout\ distribution measured by PHENIX experiment in \dau\ collision at
	 \s=200 \gevc\ \cite{Adler:2005ad}.
	 }
	\label{fig:DYplot}
\end{center}
\end{figure}
The main contribution to \kt{}\ comes from the from the Gaussian-like soft gluon showering (see Fig \ref{fig:DYplot}) and can be reproduced  relatively well 
by resummation technique \cite{Werner_resummation}. The $\mu\mu$-pair \pt{}\ distributions for the two different invariant mass bins (data taken from \cite{Ito:1980ev}) are shown on the upper panel and of Fig \ref{fig:DYplot}. It is evident that the Gaussian shape dominates over the entire range of measured \pt{\mu\mu}\ with a hint of small excess above \pt{\mu\mu}=3-4\gevc. The \pout\ measurement at higher \s\ regime reveals also similar Gaussian shape with more pronounced enhancement at large values of \pout\ indicating presence of NLO radiation. 

\begin{figure}[h]
\begin{center}
  	\resizebox{6.5cm}{!}{\includegraphics{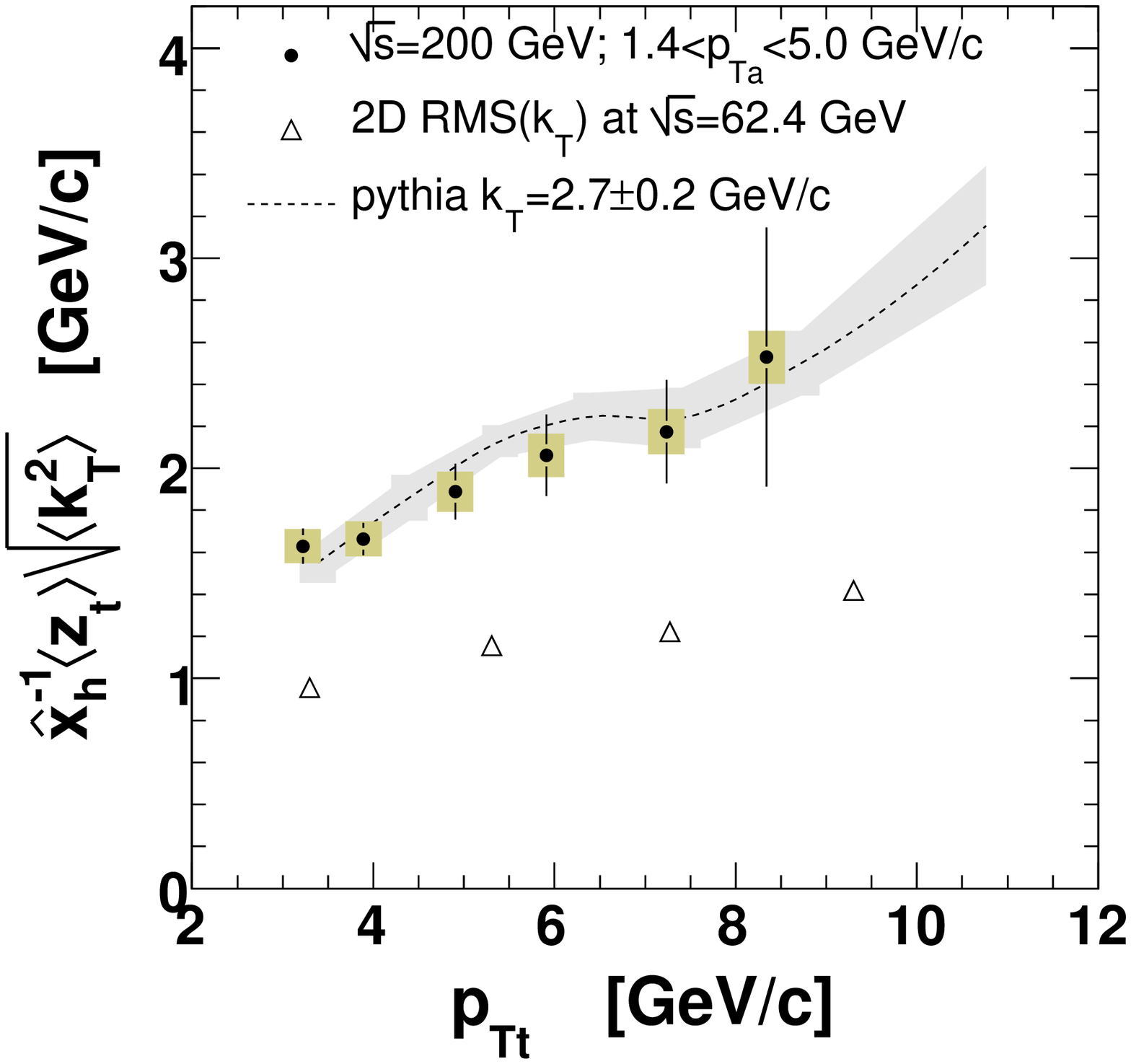}}
  	\resizebox{6.5cm}{!}{\includegraphics{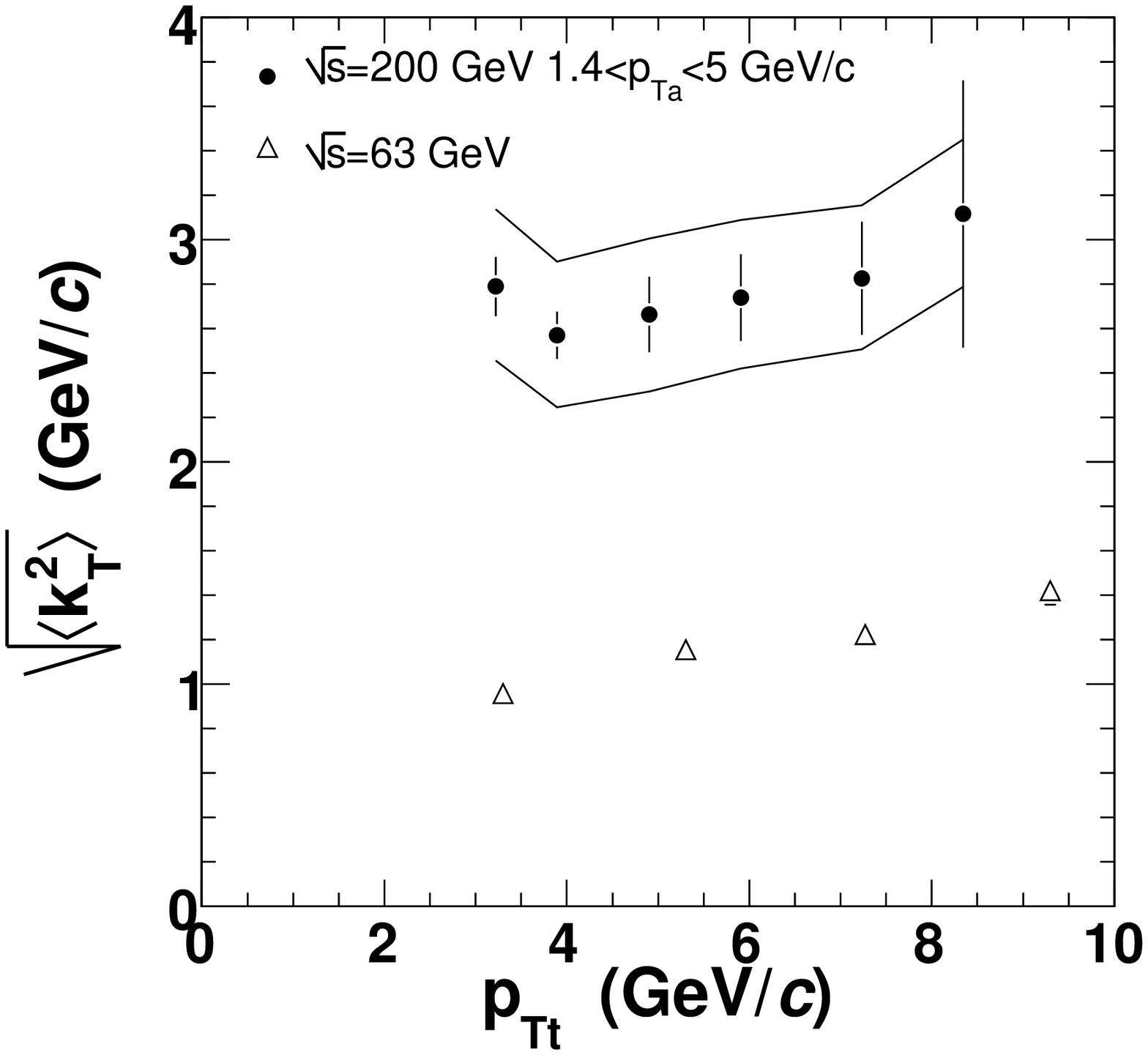}}
	\caption{ {\sl Left panel:} Measured  values of \xzkt\ (see \eq{eq:zkt}) for $1.4<\pt{a}	<5$ \gevc\ as a function of \pt{t}\  in \pp\ at \s=200 \gev. CCOR measurement of 
	\sqrtrms{\kt{}}\ values at \s=62.4 \gev\ (open triangles) \cite{CCORjt}. The solid 
	line represent the PYTHIA simulation with \sqrtrms{\kt{}}=2.7 \gevc. The gray band
	 correspond to the variation of the input \sqrtrms{\kt{}} value by $\pm$ 0.1 \gevc.
	{\sl Right panel:} 
	 Extracted \sqrtrms{\kt{}}\ values when \xhq\ and \mzt\ were evaluated
	  analytically (see \cite{longPRC:2006sc}). The solid lines represent the systematic
	  error bars originating mainly from the uncertainty on the fragmentation function. 		  CCOR data (open triangles) the same as on the left panel.
	 }
	\label{fig:zkt_kt}
\end{center}
\end{figure}

In this analysis we used the two-particle azimuthal correlation function to determine \mean{\pout}\ for different trigger (\pt{t}) and associated (\pt{a}) particle transverse momentum bins (for details see \cite{longPRC:2006sc}). We have shown that
the \sqrtrms{\pout}\ value can be related to \sqrtrms{\kt{}}\ as
\begin{equation}
\label{eq:zkt}
{\mean{\zt(\kt{},\xh)}\sqrtrms{\kt{}}\over\xhq(\kt{},\xh)} =
{1\over\xh}\sqrt{\rms{\pout}-\rms{\jt{y}}(1+\xh^2)}`
\end{equation}
where \xh=\pt{a}/\pt{t}\ and \jt{y} is the jet fragmentation transverse momentum component in the azimuthal plane. All quantities on the right-hand side of \eq{eq:zkt} can be directly extracted from the correlation function while the left-hand side corresponds to the product of\sqrtrms{\kt{}}, \xhq=\mean{\ptq{a}}/\mean{\ptq{t}}, the ratio of the mean associated to trigger parton momenta, and \mzt=\mean{\pt{t}/\ptq{t}}, the mean momentum fraction carried by the trigger particle. The \xhq\ quantity accounts for the jet momenta imbalance due to the \kt{}\ bias (\kt{}\ vector is more likely to be colinear with the trigger jet for events selected with the \pt{t}$>$\pt{a}\ condition). The \mzt\ quantity accounts for the fact that even when the trigger particle momentum \pt{t}\ is fixed, the trigger parton momentum \ptq{t}\ varies with \pt{a}, the transverse momentum of  associated particles.

Measured \xzkt\ values for different trigger particle momenta \pt{t}\ are shown on \fig{fig:zkt_kt}. 
In order to determine the \sqrtrms{\kt{}}\ values we have to evaluate the \xhq\ and \mzt\  quantities.
We assumed that

\begin{equation}
\label{eq:mzt}
\mzt\approx
\int_{x_{T\rm t}}^1 \zt^{-1}\;\D(\zt)
\int_0^{\sqrt{s}/2}\sq\int_0^\pi
\pn\,G(\pn)\cdot\D({\pt{a}\over\ptq{a}})
{1\over\ptqkv{a}}
\;d\phi\;d\ptq{}\;d\zt
\end{equation}
where \pn=\kt{t}+\kt{a}\ is a net parton-pair momentum, $G(\pn)=\exp(-\pnkv/2\rms{\kt{}})$ describes the Gaussian probability of the net pair momentum magnitude distribution, \sq\ is the unsmeared final state parton momentum distribution, \D\ is the fragmentation function.

The \xhq\ quantity can be evaluated in the similar way as \mzt\ (for details see \kt{} smearing section of \cite{longPRC:2006sc}).  As it has been demonstrated in \cite{longPRC:2006sc} it is not possible to extract the fragmentation function from di-hadron correlations, thus we used the \D\ parameters extracted from \ee\ data \cite{Delphi_Dz_EPJ00,Opal_Dz_ZPhys}. The final values of \sqrtrms{\kt{}}\ for various values of \pt{t}\ are displayed on  \fig{fig:zkt_kt}\ (solid circles). The average value obtained is
\ktfinal.

\begin{figure}[h]
\begin{center}
  	\resizebox{6.5cm}{!}{\includegraphics{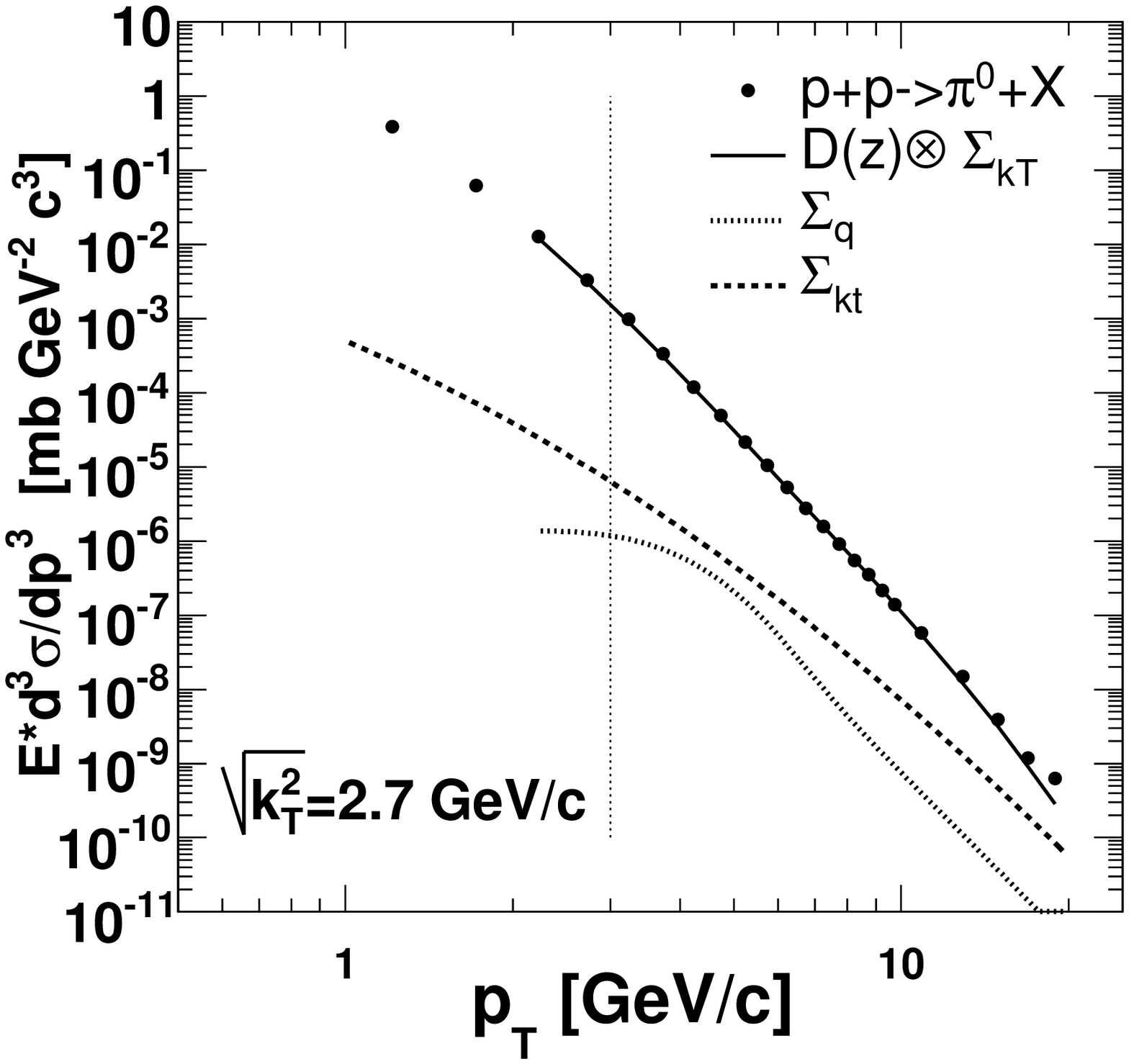}}
  	\resizebox{6.5cm}{!}{\includegraphics{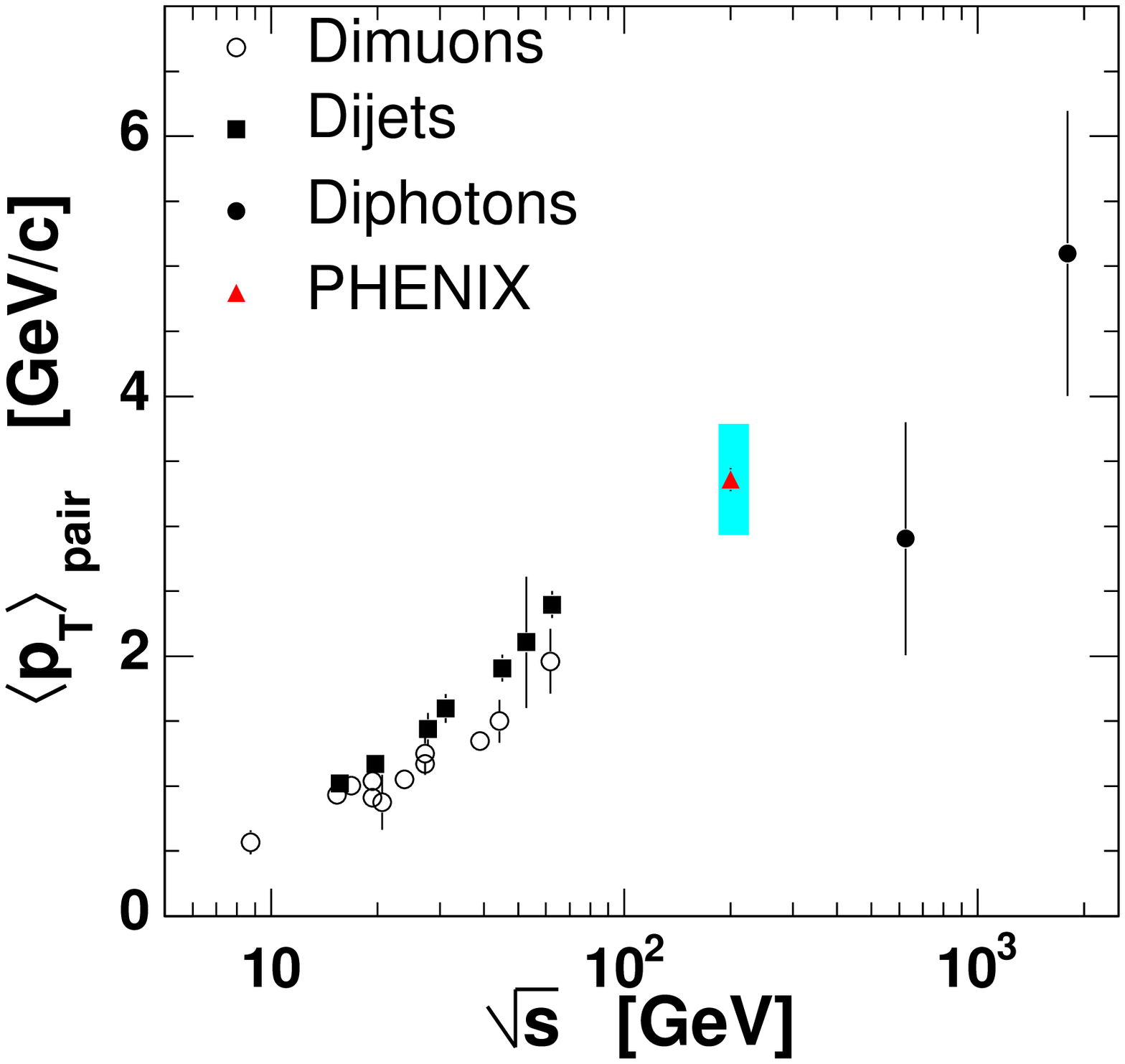}}
	\caption{{\sl Left panel:} \piz\ invariant cross section from 
	\cite{PHENIX_pi0ppPRL}.
	The dashed lined ($\Sigma_q$) corresponds to the power law parameterization of 
	the finall state parton spectra before fragmentation. The dotted line  
	($\Sigma_{kT}$) is the \kt{}-smeared $\Sigma_q$ function. The solid line
	corresponds to the folding of $\Sigma_{kT}$ with the f
	ragmentation function. {\sl Right panel:} Net parton pair momentum 
	$\mean{\pt{}}_{pair}$=$\sqrt{\pi/2}\sqrtrms{\kt{}}$ is 
	compared to compilation from \cite{Apanasevich_kt_E609}.
	 }
	\label{fig:fafit}
\end{center}
\end{figure}

We have also studied the effect of \kt{}\ smearing on the inclusive particle distributions. Although the NLO calculation without intrinsic \kt{}\ shows relatively good agreement with 
an inclusive \piz\ invariant cross section \cite{PHENIX_pi0ppPRL}, we have found that there is  a good agreement with the data when we perform a \kt{}-smearing of parametrized parton spectra by \kt{}\ magnitude 2.7 \gevc\ (see left panel of 
\fig{fig:fafit}). 

The \piz\ cross section was calculated as a folding 
$D(z)\otimes\Sigma_{kT}({\pt{}\over z})$, 
%according to 
%\begin{equation}
%{1\over\pt{}}{d\sigma_\pi\over d\pt{}} \propto \int_{xT}^1 {dz\over z^2} D(z) \Sigma_{kT}({\pt{}\over z})
%\end{equation}
where $\Sigma_{kT}(\pt{})$ is the smeared final state parton spectrum.
%is evaluated as
%\begin{equation}
%\Sigma_{kT}(\pt{}) \propto \int_0^{\sqrt{s}/2}\sq\exp\left( {(\pt{}/z-\ptq{})^2 \over\mean{\ktkv{x}}} \right)d\ptq{}
%\end{equation}
%where \mean{\ktkv{x}}=\mean{\ktkv{}}/2.  
The unsmeared ($\Sigma_q$) and \kt{}-smeared ($\Sigma_{kT}$) final state parton  spectra are shown as a dashed and dotted lines on the left panel of \fig{fig:fafit}. 
The value of the net-transverse momentum of the pair found in this analysis is shown on the right panel of \fig{fig:fafit}.  This value appears to be with a good agreement with the world  data from  \cite{Apanasevich_kt_E609}. It is interesting to note that at LHC energies the magnitude of \kt{}\ will be as large as 5-7 \gevc. 

\bibliographystyle{h-physrev3}
\bibliography{jan_HP06_proc}

\end{document}